\documentclass[letter]{aa}

\usepackage{natbib}
\usepackage{graphicx}
\usepackage{txfonts}
\usepackage[colorlinks,allcolors=blue]{hyperref}

\usepackage{multirow}
\usepackage{xspace}
\usepackage{scalerel}
\usepackage[normalem]{ulem}

\newcommand{\Zi}{Z_{\rm i}}
\newcommand{\Msun}{{\rm M}_{\odot}}
\newcommand{\Rdom}{R_{\rm dom,0}}
\newcommand{\Pdom}{P_{\rm dom,0}}
\newcommand{\Rb}{R_{\rm b}}
\newcommand{\Pb}{P_{\rm b}}
\newcommand{\Rs}{R_{\rm s}}

\newcommand{\logAge}{\log(\tau / {\rm yr})}
\newcommand{\pfm}{P_{\rm FM}}
\newcommand{\Mini}{M_{\rm i}}

\newcommand{\ks}{K_{\rm s}}
\newcommand{\wjk}{W_{\scaleto{\rm J,K_s}{4.5pt}}}
\newcommand{\wbprp}{W_{\scaleto{\rm BP,RP}{4.5pt}}}

\newcommand{\gaia}{\textit{Gaia}}
\newcommand{\simbad}{\texttt{SIMBAD}}
\newcommand{\cprime}{C$^{\prime}$}

\begin{document} 

  \title{The period-age relation of long-period variables}
  \subtitle{}
\titlerunning{The period-age relation of LPVs}
\authorrunning{Trabucchi et al.}

\author{M.~Trabucchi\inst{\ref{inst_gen}}\fnmsep\thanks{Corresponding author: M. Trabucchi (\href{mailto:michele.trabucchi@unige.ch}{\tt michele.trabucchi@unige.ch})},
N.~Mowlavi\inst{\ref{inst_gen}}
}
\institute{
Department of Astronomy, University of Geneva, Ch. Pegasi 51, 1290 Versoix, Switzerland\label{inst_gen}
}
\date{December 2021}

\abstract
{Pieces of empirical evidence suggest the existence of a period-age relation for long-period variables (LPVs). Yet, this property has hardly been studied on theoretical grounds thus far.}
{We aim to examine the period-age relation using the results from recent nonlinear pulsation calculations.}
{We combined isochrone models with theoretical periods to simulate the distribution of fundamental mode LPV pulsators, which include Miras, in the period-age plane, and we compared it with observations of LPVs in Galactic and Magellanic Clouds' clusters.}
{In agreement with observations, models predict that the fundamental mode period decreases with increasing age because of the dominant role of mass in shaping stellar structure and evolution. At a given age, the period distribution shows a non-negligible width and is skewed toward short periods, except for young C-rich stars. As a result, the period-age relations of O-rich and C-rich models are predicted to have different slopes. We derived best-fit relations describing age and initial mass as a function of the fundamental mode period for both O- and C-rich models.}
{The study confirms the power of the period-age relations to study populations of LPVs of specific types, either O-rich or C-rich, on statistical grounds. In doing so, it is recommended not to limit a study to Miras, which would make it prone to selection biases, but rather to include semi-regular variables that pulsate predominantly in the fundamental mode. The use of the relations to study individual LPVs, on the other hand, requires more care given the scatter in the period distribution predicted at any given age.}

\keywords{stars: AGB and post-AGB -- stars: evolution -- stars: variables: general -- Galaxy: stellar content -- Galaxy: globular clusters: general -- Magellanic Clouds}

\maketitle

\section{Introduction}
\label{sec:Introduction}

Low- to intermediate-mass stars approach the end of their lives through the asymptotic giant branch (AGB) evolutionary phase, during which they exhibit pulsations with timescales up to several hundreds of days, and they are hence known as long-period variables (LPVs). If their $V$-band amplitude exceeds 2.5 mag, they are classified as Miras, which have a rather regular periodicity and they are believed to pulsate only in the radial fundamental mode (FM). If their photometric amplitude is smaller, they are known as semi-regular variables (SRVs), which are thought to be the progenitors of Miras. The name stems from the lesser degree of regularity of their light curves, likely due to the fact that they can pulsate in multiple modes simultaneously.

The notion that younger LPVs tend to display longer periods compared to older ones, often referred to as the period-age (PA) relation, is rooted in the empirical evidence from stellar kinematics in the solar neighborhood. The first such piece of evidence is probably due to \citet{Merrill_1923}, who pointed out that M-type LPVs increasingly lag behind the local standard of rest (i.e., possess a higher asymmetric drift) as their period decreases. Later studies \citep[as summarized by][]{WyattCahn_1983} confirmed this behavior \citep[also using proper motion data, e.g.,][]{WilsonMerrill_1942}, and showed that the shorter periods are also accompanied by a higher velocity dispersion. Furthermore, groups of LPVs with relatively short periods are characterized by a greater scale height above the Galactic plane. This was shown, using for the first time the radial velocity of LPVs in the southern hemisphere, by \citet{Feast_1963}. In this seminal paper, Feast realized that LPVs with shorter periods must be members of older stellar populations and emphasized their highly promising applications for both Galactic and extra-galactic studies over a wide range of stellar ages. It should be noted that the PA relation is connected with the existence of a period-metallicity relation \citep[][and references therein]{LloydEvansMenzies_1973,LloydEvans_1983b,Feast_1981,FeastWhitelock_2000b}.

A number of subsequent works have corroborated the PA relation on empirical grounds, or have exploited it to interpret observational results. Relevant examples are studies of LPVs in globular clusters \citep[e.g.,][]{Feast_1966,LloydEvans_1983b,Whitelock_1986}, toward the galactic center and bulge \citep[][]{LloydEvans_1976,Feast_etal_1980,Whitelock_etal_1991} or at high galactic latitude \citep{JuraKleinmann_1992_Mira,Whitelock_etal_1994}. Of particular interest is the recent effort to extend the analysis of LPVs to dwarf galaxies in the Local Group \citep{Menzies_etal_2002,Menzies_etal_2008,Whitelock_etal_2009,Menzies_etal_2010,Menzies_etal_2011,Sakamoto_etal_2012,BattinelliDemers_2012,BattinelliDemers_2013,Whitelock_etal_2013,Menzies_etal_2015}.

The \textit{Hipparcos} mission provided the means to refine the results on the period-kinematics connection. This was done by \citet{FeastWhitelock_2000}, who found evidence supporting the existence of a bar-like structure in the Bulge from the orbits of local LPVs. A similar study dedicated to C-rich LPVs was performed by \citet{Feast_etal_2006}, who provided quantitative age estimates for these stars. A summary of the main results and prospects emerging from these \textit{Hipparcos}-era studies is given by \citet{Feast_2007}. More recently, the study of the Galaxy with LPVs has been stimulated by the wealth of data acquired by large-scale surveys \citep[e.g.,][]{Catchpole_etal_2016,Urago_etal_2020}, especially the \gaia\ mission \citep{Grady_etal_2019,Grady_etal_2020}.

It seems relevant that just a few years after the study of \citet{Feast_1963}, \citet{KippenhahnSmith_1969} predicted the PA relation of classical Cepheids from stellar evolution and pulsation models. The theoretical modeling of Cepheids and of their period-luminosity (PL) and PA relations is now an active field of research \citep[e.g.,][]{Bono_etal_2005,Anderson_etal_2016,DeSomma_etal_2020}. In contrast, when it comes to theoretical assessments of the LPV PA relation, the literature is surprisingly scarce (especially in comparison with the significant effort put into empirical studies). In fact, we were able to identify only two relevant studies addressing this subject \citep{WyattCahn_1983,Eggen_1998}. The discrepancy in period predictions between linear and nonlinear pulsation models \citep[e.g.,][]{YaAriTuchman_1996,Lebzelter_Wood_2005,Trabucchi_etal_2021}, and more generally the difficulty in modeling the structure of evolved red giants, likely played a role in hampering the theoretical investigation of the PA relation of LPVs.

Motivated by the release of updated AGB evolutionary models \citep{Pastorelli_etal_2019,Pastorelli_etal_2020} and the availability of new, accurate model predictions for the FM period of AGB stars \citep{Trabucchi_etal_2019,Trabucchi_etal_2021}, we decided to investigate the nature of the PA relation of LPVs on theoretical grounds. The adopted models and observed data are described in Sect.~\ref{sec:Methods}, while in Sect.~\ref{sec:Results} we present the results, which are discussed in Sect.~\ref{sec:Discussion}. We summarize our conclusions in Sect.~\ref{sec:Conclusions}.

\section{Methods}
\label{sec:Methods}

\subsection{Models}
\label{sec:Methods:Models}

We employed PARSEC-COLIBRI isochrones \citep{Marigo_etal_2017} with stellar evolutionary models from \citet{Pastorelli_etal_2019,Pastorelli_etal_2020} for the thermally pulsing asymptotic giant branch (TP-AGB) phase, and from PARSEC \citep[][version 1.2S]{Bressan_etal_2012} for the preceding evolution. The adopted set of isochrones covers the range 0.001 to 0.016 in initial metallicity ($\Zi$), with a 0.001 step, while it spans the age interval $8.00
\leq\logAge\leq10.45$ with a step of 0.05. Since the AGB phase is short-lived, it only spans a small range of initial masses for each given isochrone, of order of 10$^{-2}\,\Msun$ at most.

The adopted isochrones include linear pulsation periods from \citet{Trabucchi_etal_2019} for overtone modes and nonlinear periods computed with the period-mass-radius relation from \citet{Trabucchi_etal_2021} for the FM\footnote{
    Hereinafter, whenever we discuss periods, it should be understood that we refer to FM periods on which this work is focused.
}. Pulsation properties were computed along both the early-AGB and the TP-AGB. We did not extend our analysis to red supergiant stars as the pulsation prescription we employed are strictly valid only below $7\,\Msun$.

We recall that, with the adopted nonlinear relation, the period increases with radius ($R$) as a broken power law, whose exponent decreases as soon as the ``bending radius'' $\Rb$ is exceeded, it and becomes zero when the ``saturation radius'' $\Rs>\Rb$ is reached (i.e., the period becomes independent of radius). The exact values of $\Rb$ and $\Rs$, as well as of the exponents, depend on the current mass ($M$). We assume that the FM is dominant if the stellar radius is larger than the critical value $\Rdom$, which we computed from the current stellar mass using Eq.~4 of \citet{Trabucchi_etal_2021}.

\subsection{Data}
\label{sec:Methods:Data}

As a first set of data, we considered the cluster-LPV pairs used by \citet[][see their tables~1 and ~2]{Grady_etal_2019}. These consist of 19 clusters in the Large Magellanic Cloud, hosting a total of 20 potential LPV members, and eight Galactic clusters each hosting a potential LPV member.

We expanded this list with data for LPVs in a few populous clusters, namely the Galactic clusters NGC 362, NGC 2808, 47 Tuc (NGC 104), and $\omega$~Cen (NGC 5139); the LMC clusters NGC 1978 and NGC 1846; and the cluster NGC 419 in the Small Magellanic Cloud (SMC). The source lists were taken from \citet{Lebzelter_Wood_2005,Lebzelter_Wood_2007,Lebzelter_Wood_2011,Lebzelter_Wood_2016} and \citet{Kamath_etal_2010}, whose notation for the sources names is adopted here. After excluding the star LW3 in NGC 1846 and the star V129 in $\omega$ Cen, which are unlikely cluster members \citep[cf.][]{Lebzelter_Wood_2007,Lebzelter_Wood_2016}, we reached a total of 203 sources.

The aforementioned studies also provide a lot of information, possibly including $JHK$ photometry, one or more periods, and a spectral type. In order to expand on the available data, we crossmatched the selected sample with the Two Micron All-Sky Survey \citep[2MASS,][]{Skrutskie_etal_2006}, the all-sky data release of the Wide-field Infrared Survey Explorer \citep[AllWISE,][]{Cutri_etal_2013}, the catalog of variable stars from the All-Sky Automated Survey for SuperNovae \citep[ASAS-SN][]{Jayasinghe_etal_2020}, the catalogs of LPVs in the Magellanic Clouds from the third phase of the Optical Gravitational Lensing Experiment \citep[OGLE-III,][]{Soszynski_etal_2009_LMC,Soszynski_etal_2011_SMC}, the early third data release from the \gaia\ mission \citep[\gaia\ EDR3,][]{GaiaCollaboration_2021_EDR3}, and the catalog of LPV candidates from \gaia\ DR2 \citep{Mowlavi_etal_2018}.

Following \citet{Grady_etal_2019}, we took ages from \citet{Kharchenko_etal_2016} and \citet{Baumgardt_etal_2013} for clusters in the Galaxy and LMC, respectively, thereby ensuring that ages would be homogeneously derived for clusters in both galaxies. Age uncertainties from \citet{Baumgardt_etal_2013}, provided for each cluster, are generally around $\sigma_{\log(\tau)}\simeq0.05$. \citet{Kharchenko_etal_2016} do not provide age uncertainties, but a reasonable upper limit for their method should be $\sigma_{\log(\tau)}=0.2$ based on the analysis of \citet{Kharchenko_etal_2005} \citep[the same value was adopted by][in their Fig.~7]{Grady_etal_2019}.

As discussed by \citet{Kamath_etal_2010}, the age of the SMC cluster NGC 419 is believed to be around 1.4-1.6 Gyr. This is consistent with the value $\tau=1.45\pm0.05$ Gyr from \citet{Goudfrooij_etal_2014}, while it is as young as $\tau\simeq0.89\pm0.015$ Gyr according to \citet{Perren_etal_2017}. Since an accurate estimate is not necessary for our exploratory analysis, we took a rough average and assumed $\logAge=9.1\pm0.1$. NGC 419 and NGC 1846 likely exhibit TP-AGB boosting \citep{Girardi_etal_2013}. We note that some clusters show multiple stellar populations, whose age spread has been estimated in some cases \citep[e.g.,][]{Mackey_BrobyNielsen_2007,Joo_Lee_2013,Villanova_etal_2014} and is consistent with the age uncertainties we adopted.

Distances of Galactic clusters were also taken from \citet{Kharchenko_etal_2016}, while for the Magellanic Clouds and their clusters we adopted the distance moduli $\mu_{\rm LMC}=18.49\pm0.09$ mag and $\mu_{\rm SMC}=18.96\pm0.02$ mag from \citet{deGrijs_etal_2017}. We searched for data on interstellar extinction from several literature works \citep[e.g.,][]{Nayak_etal_2016,Kharchenko_etal_2016,Perren_etal_2017}, all of which suggest that extinction in the $\ks$ filter is smaller than $\sim0.1$ mag for most of the clusters we considered, and at most as large as $\sim0.3$ mag, which is negligible for our purposes.

A detailed membership verification is beyond the scope of this work, and we relied on the checks performed by authors whose source lists we adopted. It should be kept in mind that some sources may not be real cluster members.

For sources without a spectral type, we used the \gaia-2MASS diagram \citep{Lebzelter_etal_2018,Lebzelter_etal_2019} to determine whether they are O- or C-rich. We used the near-infrared period-luminosity diagram to identify the most likely pulsation mode associated with each period of each observed source. We selected only FM periods and rejected long secondary periods and periods attributed to overtone mode pulsation. The details of these classification steps are provided in Appendix~\ref{asec:ClassificationOfObservedLPVs}. Out of 203 sources from the initial list, we identified 95 LPVs pulsating in the FM, consisting of 40 C-rich and 55 O-rich sources. They consist of 29 Miras, 33 semi-regular variables, and 33 other sources (most likely LPVs) whose variability type has not been determined. We note that, with the exception of \gaia\ DR2, the sources of variability data considered here do not report the uncertainty associated with observed periods. However, since periods were derived in most cases from well-sampled, high-quality variability observations, relative period uncertainties are most likely negligible compared with those associated with age.

\begin{figure*}
    \centering
    \includegraphics[width=.9\hsize]{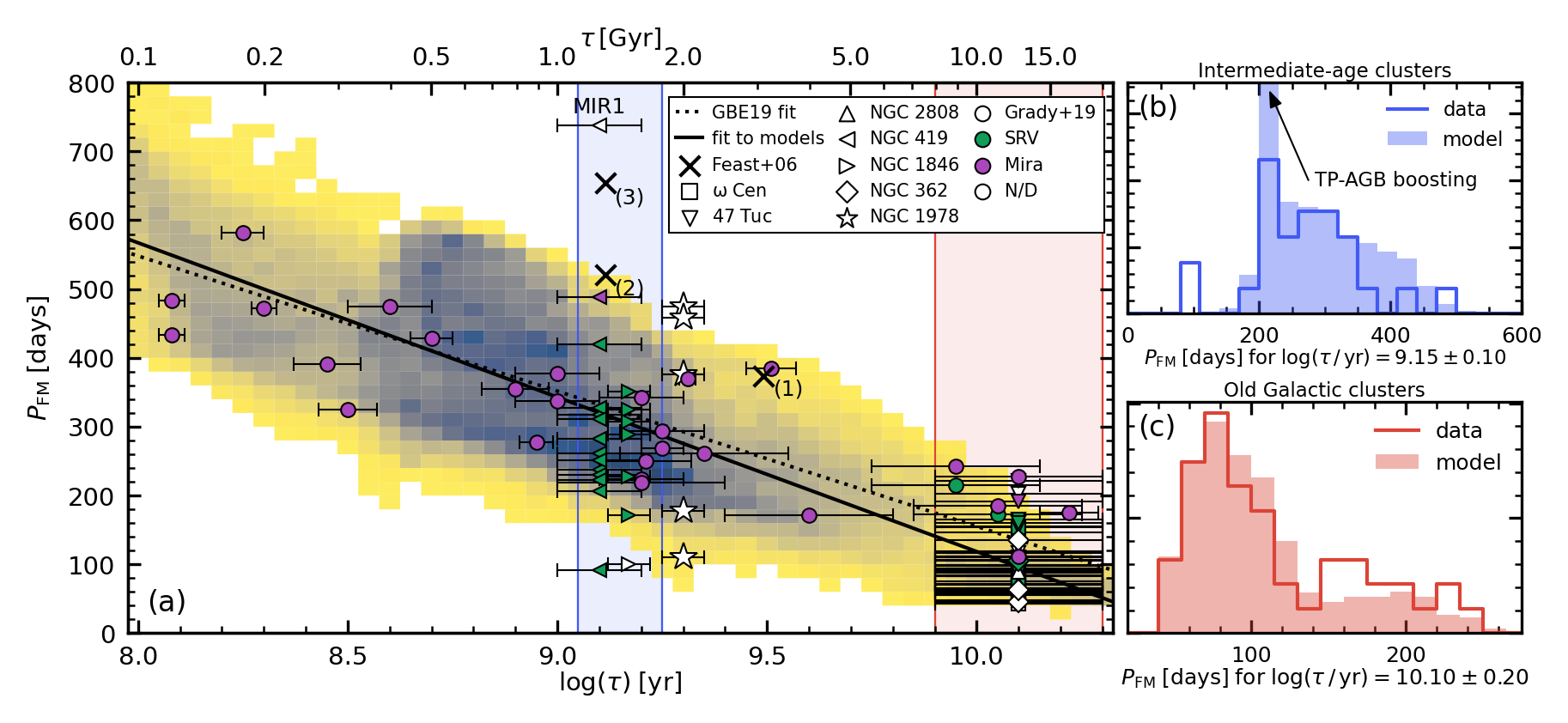}
    \caption{Period-age diagram. Panel (a) shows the predicted period-age distribution (darker tones indicate a higher expected number of LPVs on a linear scale, normalized to maximum). Symbols represent observed LPVs (green: SRVs; purple: Miras; white: unclassified) with the shape indicating their host cluster or literature source as indicated in the legend. The age uncertainties are marked by the error bars. The groups of galactic C-stars of \citet{Feast_etal_2006} are marked by crosses annotated with the group number. The solid and dotted line represent a linear best-fit to models and the best-fit by \citet{Grady_etal_2019}, respectively. Period distributions at selected ages are compared in panels (b) and (c) and marked in panel (a) by the blue and red shaded areas  (at $\logAge\sim9.15$ and $\sim10.10$, respectively). For clarity, the effect of the TP-AGB boosting is suppressed in panel (a).}
     \label{fig:Page_oldPdist}
\end{figure*}

\section{Results}
\label{sec:Results}

\begin{figure*}
    \centering
    \includegraphics[width=.9\hsize]{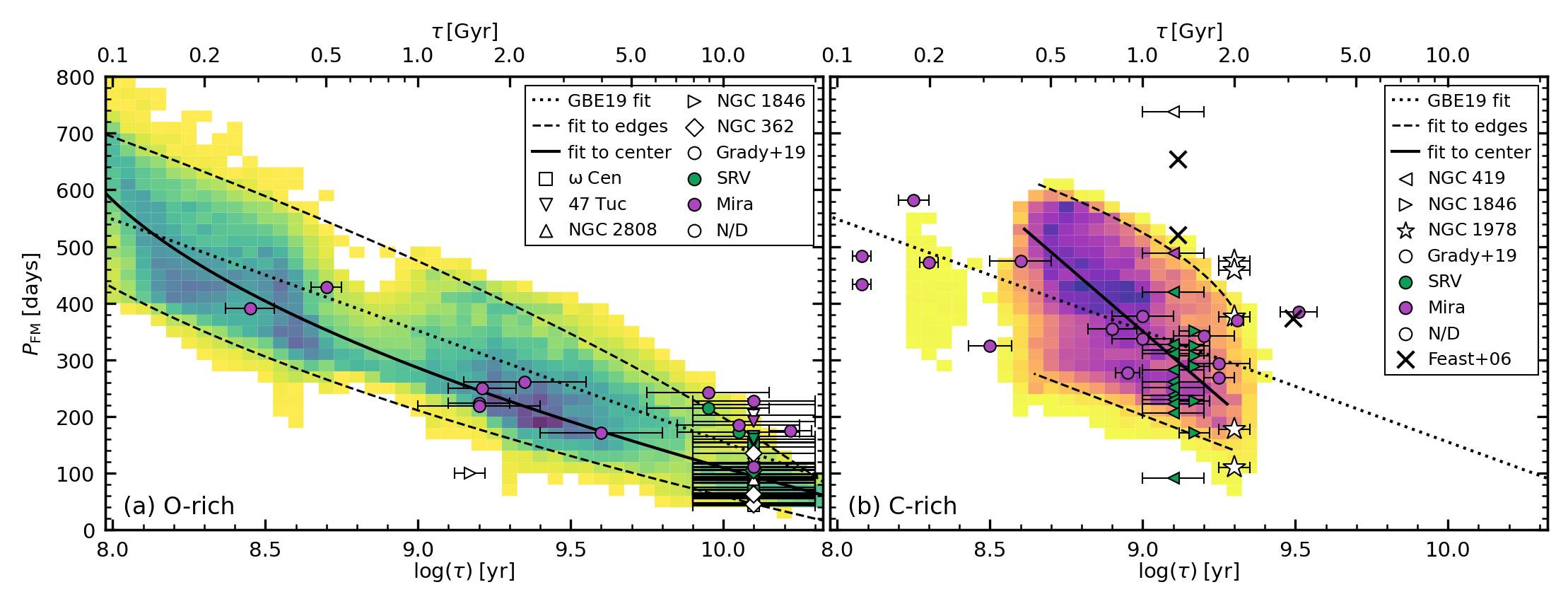}
    \caption{Similar to Fig.~\ref{fig:Page_oldPdist}, but limited to O-rich (left panel) and C-rich (right panel) LPVs. The solid line marks the best fit to the models, while dashed lines are best fits to the edges of the model distribution (see the text for more details).}
     \label{fig:Page_ChemType}
\end{figure*}

Panel (a) of Fig.~\ref{fig:Page_oldPdist} shows a comparison between model predictions and observations in the $P_{\rm FM}$--$\logAge$ plane. The former are displayed by a density map showing the expected number $N_{\rm FM}$ of LPVs pulsating in the FM in each period-age bin, normalized to maximum. Model predictions are in good agreement with data derived from observations (i.e., individual LPVs in clusters, represented by symbols), and they show that the period of LPVs pulsating in the FM decreases with increasing age. Crosses mark the average properties of the three groups of C-rich LPVs from \citet[][their table~4]{Feast_etal_2006}, which fit the general pattern with the exception of their group 3, estimated to be older than what our models predict at $P\simeq650$.

We also show a linear best-fit to the models distribution (weighted by $N_{\rm FM}$), which shows a fairly good agreement with the best-fit to observations by \citet[][also shown]{Grady_etal_2019}. However, the best-fit line does not fully capture the properties of the predictions, nor of the observed trend. Indeed, models are indicative of a substantial dispersion around the relation. For instance, at 1 Gyr, the FM period ranges from $\sim200$ days to $\sim550$ days. Conversely, LPVs pulsating in the FM with a period of 350 days are predicted to be at least $\sim$200 Myr old, but they can be as old as $\sim$3 Gyr. Observed data are consistent with the predicted spread, although the agreement cannot be considered as the observed sample adopted is not complete.

Nonetheless, it is relevant that some clusters host multiple LPVs, which are thus almost coeval, and they do span a wide period range. Some of these clusters host multiple stellar populations that are believed to have formed over a time comparable with the age uncertainties we adopted. This means that longer-period (more massive) LPVs in these clusters probably lean toward the lower age limit assumed for their host cluster, and the opposite is true at shorter periods. This tends to strengthen the agreement between models and observations.

Our data set samples the intermediate-age range (NGC 419 and NGC 1846) relatively well as well as old ages ($\omega$ Cen, 47 Tuc, NGC 362, and NGC 2808). This provides us with the opportunity to study the period distribution at these ages, and for a more detailed comparison between models and observations. On the basis of the average age of these two groups of clusters and the associated uncertainty, and taking the discrete age sampling of the isochrones into account, we considered the age ranges $\logAge=9.15\pm0.10$ and $\logAge=10.10\pm0.20$. Period distributions at those ages are displayed in panels (b) and (c) of Fig.~\ref{fig:Page_oldPdist}, respectively, showing good agreement between model predictions and observations. We note that in both cases, the distribution is skewed toward short periods, which seems to be true at all ages for O-rich stars. This can be seen in panel (a) of Fig.~\ref{fig:Page_ChemType}, which is a version of the PA plane limited to an O-rich composition\footnote{
    A further version of the PA plane highlighting both chemical types can be found in Fig.~\ref{fig:Page_ChemType_1panel} of appendix~\ref{asec:ClassificationOfObservedLPVs:SpectralType}.
}. Indeed, although at $\tau\lesssim5$ Gyr the observed sample is very scarce, it appears to be consistent with models predicting a more densely populated region in the shorter-period half of the PA distribution.

The case of C-stars, shown in panel (b) of Fig.~\ref{fig:Page_ChemType}, is different. They only form over a restricted range of initial masses and ages, so their occurrence in a given stellar population is an age indicator on its own. Toward the low-mass (old age) side of the C-star regime, the behavior is similar to the O-rich case with a concentration around relatively short periods. C-rich models tend to have a lower surface temperature and larger radii, at a given mass, compared to O-rich models, and thus they attain longer periods more easily. This occurs in particular toward higher masses, so that younger C-rich models are more concentrated at longer periods, leading to a steeper PA relation compared with the O-rich case. These predictions agree with observations on the old side of the period distribution, while the scarcity of C stars at $\tau\simeq0.6$ Gyr prevents us from performing a comparison at younger ages.

 In appendix~\ref{asec:FittingRelations}, we provide analytic PA relations by fitting the high-density parts of the O- and C-rich models' distribution. We emphasize that, because of the large scatter of the relation, ages estimated in this way for individual LPVs are bound to be highly uncertain. As a way to assess the error in age determination, we also provide analytic best-fit relations to the boundaries of the PA distribution of the models in the appendix. These relations are displayed in Fig.~\ref{fig:Page_ChemType}.

\section{Discussion}
\label{sec:Discussion}

In general agreement with observations, models confirm that LPVs pulsating predominantly in the FM follow a PA relation, which exhibits a non-negligible dispersion. Thanks to the newly available nonlinear period predictions, we were able to better examine the nature of this relation and the origin of its scatter.

The PA relation is intimately connected with the PL relation, both patterns emerging because of the prominent role of mass in shaping stellar structure and evolution. Indeed, stellar mass determines the lifetimes of the main evolutionary stages, and thus the age of stars in the AGB phase. Pulsation models \citep{Trabucchi_etal_2021} show that the radius $\Rdom$ (and corresponding luminosity) at the onset of dominant FM pulsation (DFMP) increases with mass, so that the most massive FM-dominated LPVs are brighter. They also have longer periods, as this increases with radius. In other words, the period, luminosity, and age near the tip of the AGB are all functions of initial stellar mass (at least to a good approximation).

We note that this would not be the case if the FM were dominant along the entire AGB, as the large change in radius during this phase would result in a wide range of periods at a given age. It is the very fact that DFMP occurs only during the final portion of the AGB that limits the range of periods a FM-pulsating LPV can have at a given age. Yet, the DFMP part of the AGB is long enough for significant variations in radius to occur, which result in the dispersion of the PA relation seen in Fig.~\ref{fig:Page_oldPdist}.

At a given initial metallicity $\Zi$, the shape of the period distribution primarily results from the fact that, throughout the TP-AGB (the stage during which the FM is normally excited), the envelope expansion accelerates, while the period becomes progressively less sensitive to changes in radius (see Appendix~\ref{asec:TheShapeOfThePeriodDistribution}). In particular, the slope of the period-radius relation decreases sharply at $\Pb=P(\Rb)$. The FM period distribution is roughly symmetric around that value, but at its short-period side, the FM is not dominant. Therefore, when only FM-dominated LPVs are considered, as is done here, the observed period distribution appears skewed toward short periods.

This feature is strengthened when a set of isochrones is considered which spans a range of initial metallicities because the adopted criterion for the onset of DFMP does not depend on metallicity, but the FM period does as metal-poor LPVs are warmer and have smaller radii compared with metal-rich ones. As a consequence, the bulk of the period distribution of metal-poor LPVs is at periods shorter than $\Pb$, so they only contribute to the global distribution (i.e., at all $\Zi$ at a given age) over a small period range at $P\gtrsim\Pb$. In contrast, metal-rich LPVs have periods well beyond $\Pb$, so they contribute both at that value and at longer periods. The result is an excess of FM-dominated LPVs near $\Pb$, that is to say on the short side of the overall period distribution.

We note that, in contrast with the prescription we adopted, the onset of DFMP in reality is probably sensitive to metallicity. While the good degree of agreement with observations suggests that the dependence is weak at most, it is possible for any discrepancy to be smeared out by the fact that our set of isochrone implicitly assumes a flat star-formation rate with no age-metallicity relation, so it is not an accurate representation of any realistic stellar environment. In this sense, the PA relation is environment-dependent, and it is not necessarily universal.

A further point of uncertainty stems from the fact that the prescription we adopted assumes that the FM period only depends upon the mass and radius, and that it is affected by a change in composition only through the effect that such a variation has on the radius. While this is true to a good approximation, linear models show a small dependence of periods on metallicity at a fixed mass and radius, but the quantitative impact in the nonlinear case is unknown. We can only estimate, based on the results of \citet{Trabucchi_etal_2019}, an uncertainty of $\pm10\%$ at most with respect to the prescriptions adopted here.

Qualitatively, a realistic age-metallicity relation and the metallicity dependence of the period and of the onset of DFMP are all expected to result in a steeper PA relation than the one we predict, but it is difficult to assess the relative importance of these effects.
In this sense, the composition probably affects the shape of the PA relation more than its dispersion. The latter is likely affected by the composition indirectly through mass loss, the analysis of which is beyond the scope of this study.
However, we point out that mass loss represents a source of scatter in combination with the occurrence of thermal pulses, because it reduces the minimum radius for the onset of DFMP. Thus, during the luminosity dips associated with thermal pulses, a LPV can have a period shorter than the one it had when it first entered the DFMP regime (see Appendix~\ref{asec:TheShapeOfThePeriodDistribution}). An additional source of uncertainty, which we disregarded, is rotation (or other processes that induce extra mixing in the core) which causes a spread in ages at a given initial mass \citep[cf.][for the case of classical Cepheids]{Anderson_etal_2016}.

The fairly good agreement between models and observations encourages the use of LPVs as age indicators, but the scatter of the PA relation hampers this application. We attempted to reduce the scatter through corrections involving photometric properties, as is customarily done for classical Cepheids with a color term \citep[e.g.,][]{Bono_etal_2005}, but with unsatisfactory results. A correction dependent on the photometric amplitude of variability represents a promising alternative, but it cannot be pursued at the moment. Indeed, for computational efficiency, current pulsation models include only a crude treatment of the atmospheric layers as they do not affect pulsation periods. On the other hand, the atmosphere is crucial in determining the spectral energy distribution and its variation throughout the pulsation cycle, and hence the amplitude of variability. At the same time, the observational sample adopted here is too heterogeneous for a self-consistent investigation of amplitude, but this kind of study could be made possible by the upcoming data release 3 of the \gaia\ mission \citep{GaiaCollaboration_2021_EDR3} and the future Legacy Survey of Space and Time \citep[LSST,][]{Ivezic_LSST_2019} of the Vera Rubin Observatory.

It is worth noting that our analysis applies to Miras as well as SRVs, provided that they predominantly pulsate in the FM. The limitation of PA relation studies to Miras, as has mainly been done in literature so far, undoubtedly has some advantages: to begin with, the fact that Miras are typically easier to detect than SRVs, and their light curves are easier to process as they tend to be more regular. Moreover, Miras represent the end-point of AGB evolution, so in principle they correspond to a smaller range of stellar parameters compared to the full extent of the DFMP regime, and they display a smaller range of periods at a given age \citep[cf.][]{FeastWhitelock_2000}. In other words, they should exhibit a relatively narrow PA relation (even though, based on the observational data set we adopted, there is no conclusive evidence that considering only Miras reduces the scatter of the PA relation).

Nonetheless, we caution against this approach as it is prone to introducing uncontrolled biases, as the traditional distinction between SRVs and Miras is arbitrary \citep[see][and references therein]{Trabucchi_etal_2021_SRV1}. As such, it disregards the physical processes at the origin of the range of amplitudes characterizing LPVs. In particular, photometric amplitudes are largely determined by the formation and dissociation of molecules in the stellar atmosphere, and they are likely to be metallicity-dependent. It is therefore reasonable to assume that metal-poor (old) Mira analogs might be classified as SRVs, thereby undermining the potential application of the PA relation if restricted to Miras. This seems to be supported by the fact that the bulk of old LPVs in our sample are classified as SRVs. Therefore, studies involving PA relations of LPVs would advantageously include both Miras and FM-pulsating SRVs.

The challenge associated with SRVs stems from the fact that they are often multiperiodic (even when predominantly pulsating in the FM), a property that complicates the light curve analysis and period extraction. At the same time, this feature could potentially improve age determinations as overtone modes are expected to display a PA relation as well.

\section{Conclusions}
\label{sec:Conclusions}

We used the results from recent nonlinear pulsation calculations and combined them with state-of-the-art isochrone models to investigate the PA relation of FM-dominated LPVs, finding good agreement with the distribution of observed LPVs in star clusters. The theoretical PA relation displays a non-negligible scatter, whose origin we identified due to the fact that, despite being very brief, the portion of AGB evolution during which the FM becomes dominant shows a relatively large range in mass and radius at a given age.

The theoretical distribution of FM periods is roughly symmetric, but the FM is not dominant at the shortest periods. As a result, models predict that the distribution of dominant FM periods at a given age is skewed toward short periods, in agreement with observations. Depending on stellar populations, metallicity may enhance this feature as metal-poor LPVs, which tend to be warmer and more compact, only contribute near short periods.

We provide the best-fit PA relation separately for O-rich and C-rich FM-pulsating LPVs. The latter LPVs show a steeper PA relation because of their lower surface temperatures, which allow them to reach longer periods more easily.

Our analysis concerns all LPVs predominantly pulsating in the FM, regardless of whether they are classified as Miras or SRVs. We discourage such a distinction in that it is arbitrary and prone to selection biases that risk compromising the use of LPVs as age indicators.

The main limitation in the use of the PA relation for age determinations of individual LPVs stems from its relatively large scatter. We suggest that corrective terms, involving the amplitude of variability, might help to reduce this scatter and anticipate that upcoming data from ongoing and future surveys dedicated to time-domain astronomy will be highly valuable to probe this possibility. A study of the impact of metallicity on nonlinear pulsation is highly desirable to pursue this line of investigation, as would be a theoretical investigation of the dependence of photometric amplitudes upon global stellar parameters.

\begin{acknowledgements}
M.T. and N.M. acknowledge the support provided by the Swiss National Science Foundation through grant Nr. 188697.
We are grateful to the anonymous referee for the constructive comments that helped improving this paper, and to Léo Girardi for helping with the computation and interpretation of isochrones.
This research has made use of: data from the \mbox{OGLE-III} Catalog of Variable Stars; data products from the Two Micron All Sky Survey, which is a joint project of the University of Massachusetts and the Infrared Processing and Analysis Center/California Institute of Technology, funded by the National Aeronautics and Space Administration and the National Science Foundation; data from the European Space Agency (ESA) mission {\it Gaia} (\url{https://www.cosmos.esa.int/gaia}), processed by the {\it Gaia} Data Processing and Analysis Consortium (DPAC, \url{https://www.cosmos.esa.int/web/gaia/dpac/consortium}). Funding for the DPAC has been provided by national institutions, in particular the institutions participating in the {\it Gaia} Multilateral Agreement.
This research has made use of the following free/open source software and/or libraries: the Starlink Tables Infrastructure Library \citep[STILTS and Topcat,][]{Taylor_2006}; IPython \citep{ipython} and Jupyter \citep{jupyter} notebooks; the \textsc{Python} libraries \textsc{NumPy} \citep{numpy2020}, \textsc{SciPy} \citep{SciPy}, \textsc{matplotlib} \citep[a \textsc{Python} library for publication quality graphics,][]{matplotlib}, and \textsc{Astropy} \citep[a community-developed core \textsc{Python} package for Astronomy,][]{astropy2018}. This research has made use of NASA's Astrophysics Data System Bibliographic Services, and of the following services provided by CDS, Strasbourg: the SIMBAD data base, VizieR catalogue access tool \citep[DOI: 10.26093/cds/vizier,][]{Ochsenbein_etal_2000}, the ``Aladin sky atlas'' \citep{Bonnarel_etal_2000}, and the cross-match service \citep{Boch_etal_2012,Pineau_etal_2020}.
\end{acknowledgements}

\bibliographystyle{aa}
\bibliography{references}

\appendix

\section{Classification of observed LPVs}
\label{asec:ClassificationOfObservedLPVs}

\begin{figure}
    \centering
    \includegraphics[width=\hsize]{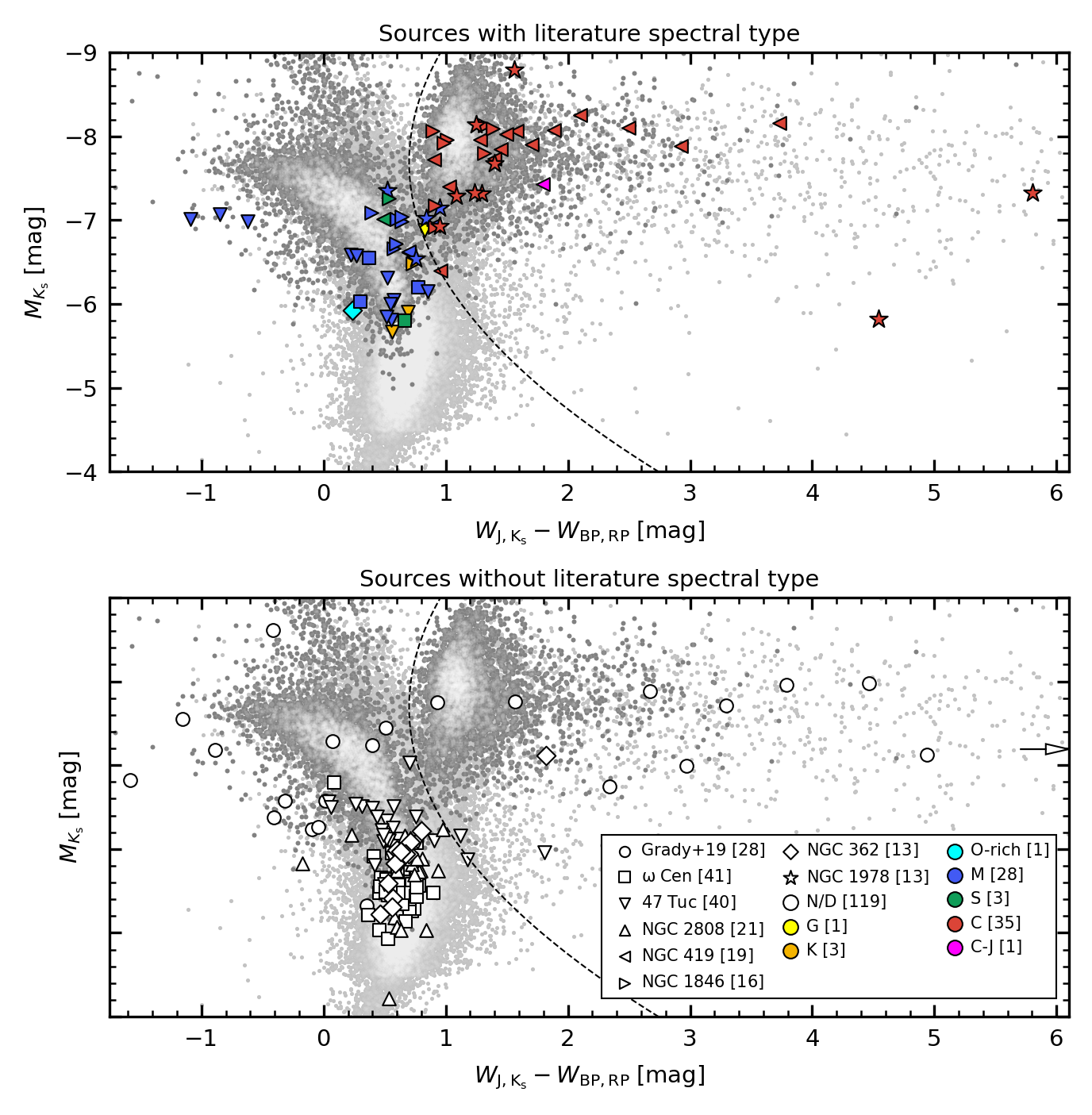}
    \caption{Absolute-$\ks$ \gaia-2MASS diagram for the stars with or without a spectral type (left and right panels, respectively) in the selected sample. Symbol colors and shapes indicate the spectral type and host cluster described in the legend, respectively, which also reports the number of sources displayed (i.e., having both optical and NIR photometry). The dashed line marks the separation between O- and C-rich sources according to \citet{Lebzelter_etal_2018}. An arrow marks the source MSX LMC 124 in NGC 1830 that, having $\wbprp-\wjk=9.73$ mag, lies outside the plot area. Background dots are LPVs in the LMC from OGLE-III (light gray) and \citet{Mowlavi_etal_2018} (darker gray).}
     \label{fig:g2md}
\end{figure}

\begin{figure*}
    \centering
    \includegraphics[width=.9\hsize]{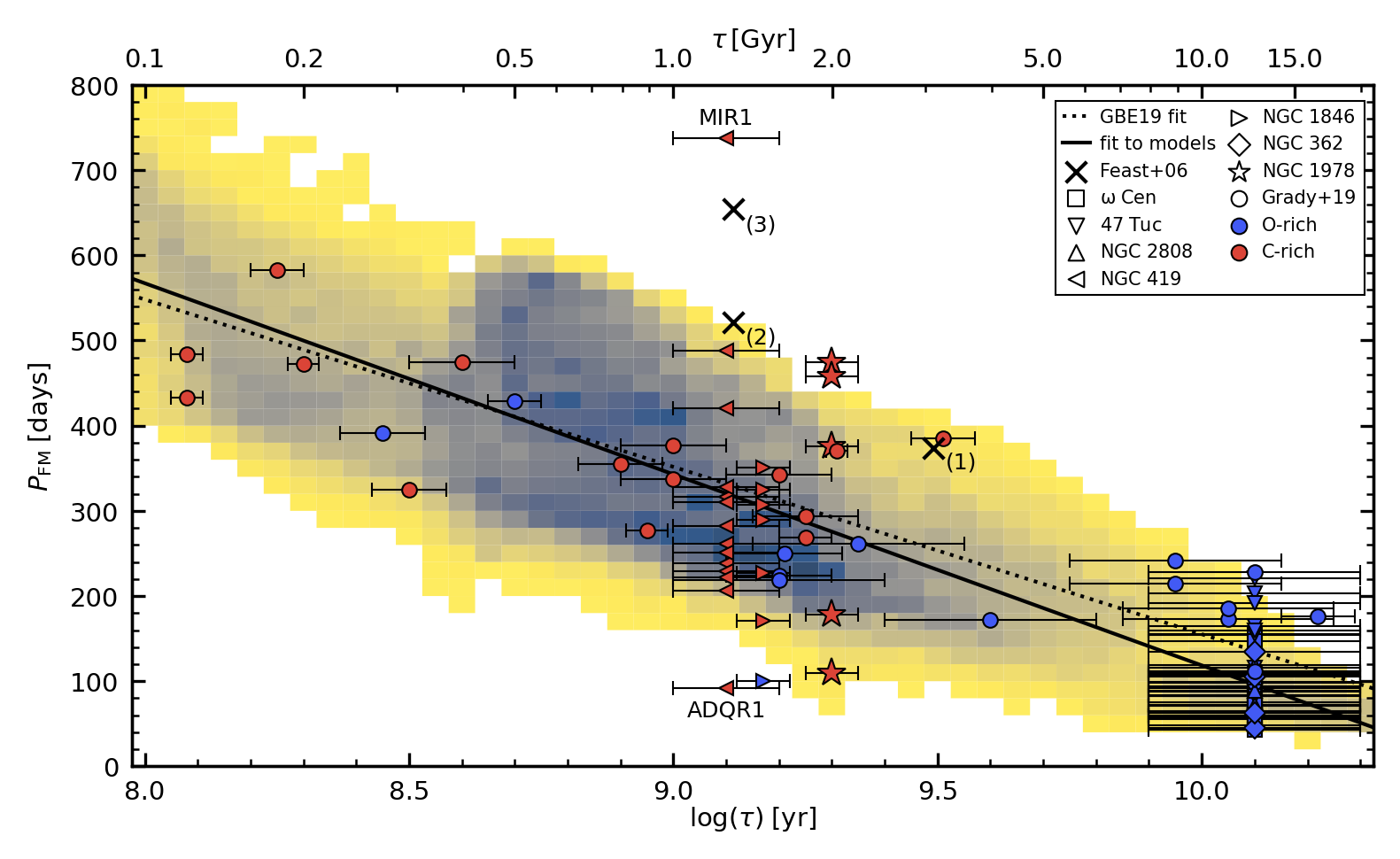}
    \caption{Similar to Fig.~\ref{fig:Page_oldPdist}, except each source is color-coded according to whether it has been classified as O-rich (blue) or C-rich (red).}
     \label{fig:Page_ChemType_1panel}
\end{figure*}

\subsection{Spectral type}
\label{asec:ClassificationOfObservedLPVs:SpectralType}

We adopted the spectral types provided by \citet{Lebzelter_Wood_2007} and \citet{Kamath_etal_2010} for 52 of the LPVs they studied in NGC 1846, NGC 1978, and NGC 419. The only exception is the star 5-3 in NGC 419, for which we adopted the S-type as reported by \citet{LloydEvans_1983}.

We also searched the \texttt{SIMBAD} astronomical database \citep{Wenger_etal_2000} for spectral type information, which we found for 26 more stars. We used the \gaia-2MASS diagram of \citet{Lebzelter_etal_2018} to confirm the chemical type classification taken from literature and to characterize the surface chemistry of sources of an unknown spectral type (see Fig.~\ref{fig:g2md}). Among the latter, we identified 13 C-rich stars and 106 O-rich sources.

Three of the sources without a spectral type lack \gaia\ photometry, so they cannot be classified with the \gaia-2MASS. Two of them (LW5 and LW22 in 47 Tuc) have no match in \gaia\ EDR3, but they have NIR data and are probably O-rich based on their position in the $J-\ks$ versus $\ks$ color-magnitude diagram. The third source is one of the two stars in NGC 1903 from the list of \citet{Grady_etal_2019}, which we identified with the 2MASS source J05171633-6920298. It is likely C-rich according to the NIR color-magnitude diagram.

Finally, the sources V138 in $\omega$ Cen, LW15 in NGC 2808, and LW4 in NGC 362 lack NIR data. They cannot be placed in the NIR PL diagram, upon which we relied to assign pulsation modes to periods, so we excluded them from the sample. The distribution of O- and C-rich sources in the period-age diagram is shown in Fig.~\ref{fig:Page_ChemType_1panel}.

\subsection{Variability}
\label{asec:ClassificationOfObservedLPVs:Variability}

For variability information, we complemented the data from Lebzelter \& Wood and \citet{Kamath_etal_2010} with the catalogs from OGLE-III, ASAS-SN, and \gaia\ DR2. Combining these data sets, we found at least one period for each of the 176 sources in our sample.

In order to identify the pulsation mode most likely responsible for periods in a given source, we assumed that the second overtone mode is associated with sequence A, the first overtone mode with sequences B and \cprime, and the fundamental mode with sequence C \citep[e.g.,][]{Trabucchi_etal_2017}. We excluded long secondary periods on sequence D as they are not due to stellar pulsation \citep[][and references therein]{Soszynski_etal_2021}, and we used the pattern of PL sequences in the LMC as a reference to guide the mode identification \citep[cf.][]{Trabucchi_etal_2021_SRV1}.

We performed this classification separately for periods coming from each distinct data set. If two or more periods from different data sets were assigned to the same pulsation mode, we retained only one of those periods, with priority to the values from Lebzelter \& Wood and \citet{Kamath_etal_2010}. If the latter authors do not provide this information, we adopted the period from OGLE-III if available, and otherwise from ASAS-SN or from \gaia\ DR2.

For some sources, the periods reported in different catalogs were assigned to the same mode through this procedure. In most cases, these periods are reasonably similar to each other. Only in a few cases were they significantly different, but this did not alter our conclusions.

When available, the variability type was taken from OGLE-III or ASAS-SN. We note that we are only interested in whether a star is classified as a Mira or semi-regular variable. In many cases, this type is not given or the star is simply considered, for instance, as an LPV or AGB in \simbad, in which case we considered the variability type as undetermined.

\section{Fitting relations}
\label{asec:FittingRelations}

We obtained analytic expressions for the PA relations separately for O- and C-rich stars, proceeding as follows. For each bin of $\logAge,$ we modeled the period distribution with a Gaussian kernel density estimator (KDE) and identified the peak of the distribution. To describe the boundaries of the PA relation, we adopted, at each age, the values of the period at which the distribution equals 25\% of its maximum. We selected this arbitrary value upon visual inspection of the PA plane. We modeled the central trend of the PA relation, as well as its short- and long-period edges, with linear or quadratic functions in the form
\begin{equation}\label{eq:pafit}
    \logAge = a_0 + a_1\,(P/\tilde{P})+a_2\,(P/\tilde{P})^2\,,
\end{equation}
(where $\tilde{P}=350$ days) and employed a Lenvenberg-Marquardt nonlinear regression algorithm\footnote{
    We made use of the \texttt{Python} library \texttt{SciPy} to perform Gaussian KDE modeling and best-fit, respectively, by means of the \texttt{gaussian\_kde} tool from the \texttt{stats} module and the \texttt{curve\_fit} function from the \texttt{optimize} module.
} to derive the best-fit coefficients, which are listed in Table~\ref{tab:pacoef}. We remark that these best-fit expressions are only valid in the intervals $8.0\leq\logAge\leq10.3$ and $20<P/{\rm days}<700$ for O-rich composition, and within $8.6\leq\logAge\leq9.3$ and $140<P/{\rm days}<620$ in the C-rich case.

\begin{figure*}
    \centering
    \includegraphics[width=.9\hsize]{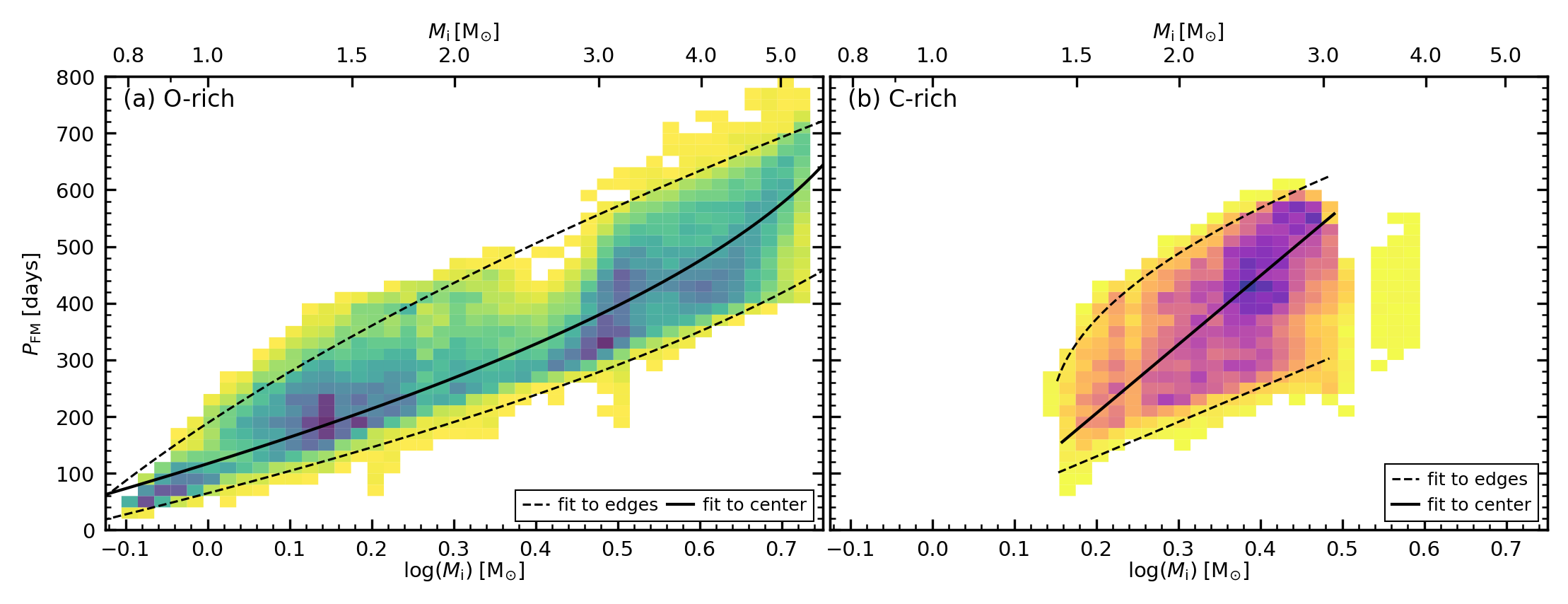}
    \caption{Similar to Fig.~\ref{fig:Page_ChemType}, but showing initial mass $\Mini$ in place of age. The best-fit lines to the most populated band and edges of the theoretical $\pfm$~--~$\Mini$ relation are shown.}
     \label{fig:PMi_ChemType}
\end{figure*}

Because of the connection between age and initial mass, the PA relation can be translated into a period-initial mass relation, which we derived using the same approach described above, and assuming the form
\begin{equation}\label{eq:pmifit}
    \log(\Mini/\Msun) = b_0 + b_1\,(P/\tilde{P})+b_2\,(P/\tilde{P})^2\,.
\end{equation}
The resulting best-fit lines are displayed in Fig.~\ref{fig:PMi_ChemType}, and the coefficients are given in Table~\ref{tab:pmicoef}.

We remark that both the PA and the period-initial mass relations depend on model assumptions, in particular mass loss and mixing, as well as on the properties of the population of LPVs, namely the star-formation history and age-metallicity relation.

\begin{table}
\caption{Best-fit coefficients for the PA relation and its boundaries in the form given in Eq.~\ref{eq:pafit}.}
\label{tab:pacoef}
\centering
\begin{tabular}{ccccc}
Sp. type & relation & $a_0$ & $a_1$ & $a_2$ \\
\hline
\multirow{3}{*}{O-rich} & center     & 10.78 & -2.660  &  0.5953 \\
                        & lower edge & 10.46 & -2.818  &  0.6578 \\
                        & upper edge & 10.54 & -0.8187 & -0.2335 \\
\hline
\multirow{3}{*}{C-rich} & center     & 9.755 & -0.7532 &         \\
                        & lower edge & 9.982 & -1.698  &         \\
                        & upper edge & 8.498 & -1.827  & -0.9959 \\
\end{tabular}
\end{table}

\begin{table}
\caption{Best-fit coefficients for the period-initial mass relation and its boundaries in the form given in Eq.~\ref{eq:pmifit}.}
\label{tab:pmicoef}
\centering
\begin{tabular}{ccccc}
Sp. type & relation & $b_0$ & $b_1$ & $b_2$ \\
\hline
\multirow{3}{*}{O-rich} & center     & -0.2790 &  0.8958 & -0.1828 \\
                        & lower edge & -0.1772 &  0.9975 & -0.2203 \\
                        & upper edge & -0.1740 &  0.2783 &  0.8247 \\
\hline
\multirow{3}{*}{C-rich} & center     & -0.0304 &  0.2885 &         \\
                        & lower edge & -0.0131 &  0.5752 &         \\
                        & upper edge & -0.2245 & -0.2720 &  0.2343 \\
\end{tabular}
\end{table}

\section{The shape of the period distribution}
\label{asec:TheShapeOfThePeriodDistribution}

\begin{figure}
    \centering
    \includegraphics[width=\hsize]{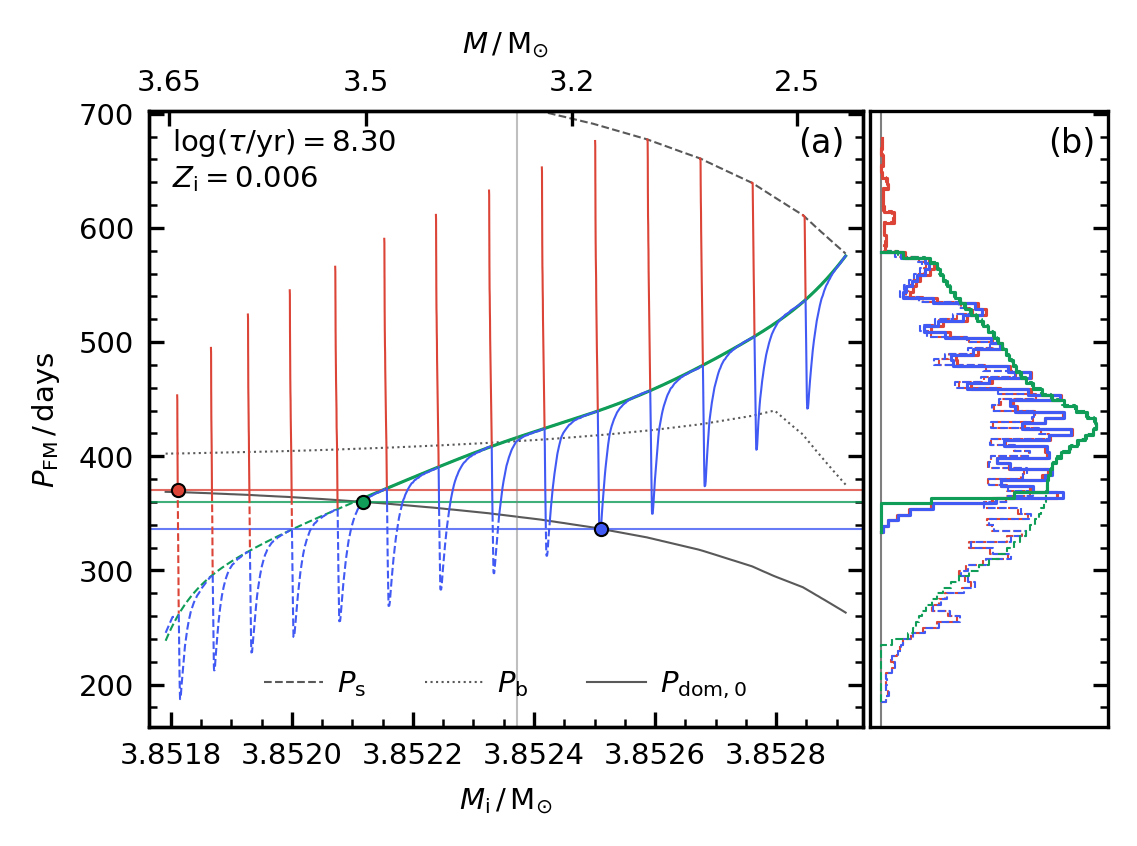}
    \caption{Period distribution at fixed age and metallicity. Panel (a) shows period as a function of initial mass (current mass on the top axis) on the TP-AGB for a $\sim200$ Myr old isochrone with $\Zi=0.006$. Red lines show full thermal pulses, while blue lines ignore luminosity spikes and green lines show only the quiescent evolution. The same color code is used for the period distributions (normalized to their maximum) on panel (b). Solid lines indicate that the FM is dominant. Circles indicate the earliest onset of DFMP accounting for (red) or ignoring (green) luminosity spikes, and the shortest period of the dominant FM (blue). Gray lines mark the critical values of periods at which the FM becomes dominant (solid line), less sensitive to radius (dotted line, which occurs at the vertical line for this specific isochrone), and independent of radius (dashed line).}
     \label{fig:Pdist_isoc_TPs}
\end{figure}

As an example case, we consider an isochrone of age $\logAge=8.3$ and initial metallicity $\Zi=0.006$. Stars on the TP-AGB have initial masses $\Mini\simeq3.85\,\Msun$ over a small range of $\sim10^{-3}\,\Msun$. The relation between period and initial mass is displayed in panel (a) of Fig.~\ref{fig:Pdist_isoc_TPs}, where isochrone portions undergoing DFMP are indicated by solid lines. Panel (b) shows the period distributions for a few different cases.

It is instructive, to begin with, to ignore the effect of thermal pulses and consider only the quiescent evolution (green lines in Fig.~\ref{fig:Pdist_isoc_TPs}). The smallest initial mass corresponds to a star that just entered the TP-AGB, when the FM has a period of $\sim240$ days but is not dominant. It only becomes dominant above a threshold radius $\Rdom$, that is for periods longer than a (mass-dependent) critical period $\Pdom$ (the solid gray line in Fig.~\ref{fig:Pdist_isoc_TPs}). The least evolved (quiescent) model with dominant FM has $\pfm\simeq360$ days (green circle and horizontal line), corresponding to a sharp cut in the period distribution shown in panel (b) of Fig.~\ref{fig:Pdist_isoc_TPs}.

As a star evolves along the AGB it expands, and its period becomes longer in response to the increase in radius. Models with a higher initial mass are more evolved, hence they have a larger radius and a longer period. The rate at which a period increases with radius is not fixed, but rather decreases with evolution. According to the prescription of \citet{Trabucchi_etal_2021}, a period grows with radius as a broken power-law with exponent $\alpha\simeq1.8$ if $R<\Rb$,  and with $\alpha\simeq1.25$ at larger radii.

This is equivalent to saying that the period grows more slowly after it exceeds a critical value $\Pb=P(\Rb)$, marked by the gray dotted line in Fig.~\ref{fig:Pdist_isoc_TPs}. The isochrone reaches it at $\Mini\simeq3.8524\,\Msun$ (vertical gray line), when $\pfm\simeq420$ days. In models with a smaller initial mass, the period is still increasing at a relatively large rate as the envelope expands, while in more massive models the period has already become less sensitive to changes in radius. This is reflected by a slight inflection of the green curve, which corresponds to the maximum in the period distribution shown in panel (b) of Fig.~\ref{fig:Pdist_isoc_TPs}. The period distribution of the full TP-AGB range is roughly symmetric around this maximum, while limiting the selection to DFMP, produces a distribution skewed toward short periods, as found in Sect.~\ref{sec:Results}.

If the luminosity dips following thermal pulses are taken into account (blue lines), the corresponding envelope contraction causes the period to decrease, and the cut at $\sim360$ days becomes less sharp. Because of mass loss, the threshold period $\Pdom$ is lowered, so that the shortest period associated with DFMP does not correspond to the least evolved model (green circle), but rather to the luminosity dip of a thermal pulse (blue circle).

To be precise, the earliest occurrence of DFMP is on the left-most luminosity spike (red circle), whose duration is so short that it is unlikely to be observed. Indeed, the inclusion of luminosity spikes alters the period distribution at long periods very little. Luminosity spikes are relevant only for relatively massive and young TP-AGB stars, and they give rise to the poorly populated portion of the PA relation at the longest periods, as seen in panel (a) of Fig.~\ref{fig:Page_ChemType}.

\end{document}